\documentstyle[12pt,fleqn]{article}
\begin{document}
\baselineskip=7.5mm

\begin{center}
{\large\bf Is GRO J1744-28 a Strange Star?}

K.S. Cheng$^{*,1}$, \,\, Z.G. Dai$^2$, \,\, D.M. Wei$^3$, \,\, and T. Lu$^{2,4}$

$^1${\em Department of Physics, University of Hong Kong, Hong Kong, China}

$^2${\em Department of Astronomy, Nanjing University, Nanjing 210093, 
China}                                                                  
$^3${\em Purple Mountain Observatory, Nanjing 210008, China}

$^4${\em CCAST (World Laboratory), P.O. Box 8730, Beijing 100080, China}

$^*${\em To whom correspondence should be addressed}

\end{center}

\begin{center}
{\bf Abstract}
\end{center}

The unusal hard x-ray burster GRO J1744-28 recently discovered by the 
Compton Gamma-ray Observatory (GRO) can be modeled as a strange star
with a dipolar magnetic field $\le 10^{11}\,$Gauss. When the accreted mass 
of the star exceeds some critical mass, its crust may break, resulting in
conversion of the accreted matter into strange matter and release of
energy. Subsequently, a fireball may form and  expand 
relativistically outward. The expanding fireball may interact with 
the surrounding interstellar medium, causing its kinetic energy to be 
radiated in shock waves, producing a burst of x-ray radiation.  
The burst energy, duration, interval and spectrum derived from such a model 
are consistent with the observations of GRO J1744-28.

\noindent
PACS numbers: 97.80.Jp; 13.38Mh; 95.30Gv; 97.60Gb

GRO J1744-28 is a new type of x-ray transient source, which was discovered
on 2 December 1995 by the Burst And Transient Source Experiment (BATSE)
on the {\em Compton Gamma-Ray Observatory} ({\it{1}}). The bursts were detected
up to energies of $\sim 75\,$keV with intervals between bursts of about
200 s initially. After 2 days the burst rate dropped to about one per 
hour ({\it{2}}). However, by 15 January 1996 the burst rate had increased to 
$\sim 40$ per day. The burst durations lasted $\sim 10\,$s.
The burst fluences (25-60\,keV) range between $1.7\times 10^{-7}$ and
$6.8\times 10^{-7}\,{\rm erg}\,{\rm cm}^{-2}$; the average fluence 
$\bar S=(2.7\pm 0.9)\times 10^{-7}\,{\rm erg}\,{\rm cm}^{-2}$. 
The position of the source is at the Galactic Center. For a distance of 
$\sim 7.5\,$kpc, the average peak luminosity was $\sim 2\times 10^{38}\,
{\rm erg}\,{\rm s}^{-1}$. Analysis of the BATSE data indicated that 
the source is a binary pulsar with a pulsation period of 0.467\,s, a companion of mass 
between 0.22--1.0\,$M_\odot$, and a binary orbital period of 11.8 days {(\it{3})}.
Because the x-ray mass function is small ($\sim 1.31 
\times 10^{-4}M_\odot$), the system must be nearly face-on to an observer
from Earth with an
inclination angle of $\sim 18^{\rm o}$ ({\it{3-5}}). Furthermore, in order for the
measured rotation period derivative to be consistent with the standard accretion
torque theory ({\it{6}}), the persistent luminosity of the source at its peak
should be close to the Eddington limit ({\it{4}}) and the surface dipole magnetic 
field of the pulsar should be $\le 10^{11}\,$G ({\it{3-5}}). 
>From the observed pulsed fraction and pulsar's x-ray spectrum,
the strength of the local surface magnetic field is estimated to be
several $10^{12}\,$G({\it{4}}). In addition, 
the proportional counter array (PCA)
experiment (2-60\,keV) on the {\em Rossi  X-ray Timing Explorer}
({\em RXTE}) ({\it{7}}) had detected GRO J1744-28 in the period 18 January 
--10 May 1996. The observations showed that following the earlier large 
bursts the flux dipped below the preburst level by up to 25\%--30\% 
and then made a slow quasi-exponential recovery back toward the preburst 
level. The observed recovery period lasted up to $\sim 1000\,$s for
some bursts, but 
most bursts recovered in a few hundred seconds. 

The properties of the hard x-ray bursts (HXRBs) from GRO J1744-28
differ from those of other known high-energy bursts --- x-ray 
bursts, soft $\gamma$-ray bursts and $\gamma$-ray bursts. First, 
the HXRBs are probably not type-I x-ray bursts 
({\it{8}}). Thus, thermonuclear flashes in matter accreted 
onto the surface of a neutron star may not produce
HXRBs. Second, the durations of the HXRBs are several hundred 
longer times than those of the three soft $\gamma$-ray repeaters even though
these two kinds have similar repeat times and spectra. Third, the HXRBs
are different from $\gamma$-ray bursts, because $\gamma$-ray bursts do not
have fast repeat times
and their spectra are much harder. On the other hand,
the repeat times and spectra of the HXRBs are somewhat similar to those 
of type-II
x-ray bursts from the Rapid Burster ({\it{2,8}}). This suggests 
that some accretion instability may be a mechanism for producing HXRBs.
Very recently Cannizzo ({\it{9}}) studied the global, time-dependent evolution 
of the Lightman-Eardley instability, which might account for some
observational features of the HXRBs.
Here we propose an alternative  model, in which a strange 
star accretes matter from its low-mass companion. 

Strange matter, which is bulk quark matter, was conjectured to be
more stable than hadronic matter ({\it{10}}). Studies  showed
that the existence of strange matter is allowable within uncertainties
inherent in a strong-interaction calculation ({\it{11}}). Thus, 
strange stars may exist in the Universe. Strange 
stars only have crusts with masses 
$\sim 10^{-5}M_\odot$({\it{12}}).  
However, the postglitch behavior of pulsars can be described by the 
neutron-superfluid vortex creep theory ({\it{13}}) which requires the
crustal mass $\ge 10^{-3}M_\odot$.
The conversion of a neutron star to a strange star may require 
the formation of a strange-matter seed, which is produced through
the deconfinement of neutron matter at a density $\sim\,$7--9$\rho_0$  
(where $\rho_0$ is the nuclear matter density), much larger than the central
density of a $1.4M_\odot$ neutron star with a moderately stiff to stiff 
equation of state ({\it{14}}). These two features suggest that strange 
stars may be formed in low-mass x-ray binaries ({\it{15}}) because when 
the neutron star in a low-mass x-ray binary accretes sufficient mass 
(perhaps $\ge 0.4M_\odot$) its central density can reach the 
deconfinement density and subsequently the whole star will undergo a 
phase transition to become a strange star. 
The phase transition from nuclear matter
to strange matter under the condition of conserved charge rather than
constant pressure may occur at a density as low as 2--3$\rho_0$({\it{16}}). 
If so, strange stars may be formed during the evolution of protoneutron 
stars ({\it{17}}). Here we would remind the reader of some arguments against
the existence of strange stars. For example, the disruption of a single
strange star may contaminate the entire galaxy and essentially all 
neutron stars may be strange stars ({\it{18}}). In view of these uncertainties, 
we should only regard strange stars as possible stellar objects.

According to the standard accretion torque theory ({\it{6}})and
Daumerie et al. ({\it{4}})
the binary system of GRO J1744-28 is nearing the end of the mass transfer 
phase. If this is true then the companion has transferred mass of 
$\ge 0.4M_\odot$ to the pulsar
and now has a mass $\sim 0.22$--$0.5M_\odot$. From the scenario proposed 
in ({\it{15}}) and the above discussion, we suggest that the pulsar in GRO J1744-28 
is a strange star.

We next discuss the burst mechanism. We assume that far from the proposed
strange star
the magnetic field is purely dipolar, and the accretion flow
is spherically symmetric. We consider a simple case: $M_{\rm pulsar}=
1.8M_\odot$ and $R_{\rm pulsar}=10^6\,$cm. The Alfv\'en radius for spherical
accretion, obtained by balancing accretion and magnetic pressure, is
given by
$R_{A}=1.8\times 10^7 L_*^{-2/7} B_*^{4/7}\,\,{\rm cm}$,
where $L_*$ is the total accretion luminosity of
$\sim 2\times 10^{38}\,{\rm ergs}\,{\rm s}^{-1}$ (the Eddington limit), 
and $B_*$ is the surface dipolar magnetic field strength in units of
$2\times 10^{10}\,$G which is close to the $B_*$ derived in ({\it{5}}). 
For an assumed dipolar magnetic field geometry,  
$\sin ^2\theta_m/r={\rm constant}$, where $\theta_m$ is the magnetic
colatitude. Thus, at the stellar surface near a pole, the cross-sectional
area of the accretion column is about
$A_p\approx 1.8\times 10^{11} L_*^{2/7} B_*^{-4/7}\,\,{\rm cm}^2$,
and the corresponding radius of the accretion column is
$r_p \approx 2.4\times 10^5 L_*^{1/7} B_*^{-2/7}\,\,{\rm cm}$.

As the strange star accretes matter from its companion, strong
pressure is formed
at the base of the accreted matter near 
the polar cap, due to the gravitational attraction of the
compact strange star. When this pressure
exceeds the critical stress of the star's thin crust, the crust may break. 
Therefore, the condition under which a crust-breaking event takes place 
should be given by
$\rho h g=\mu\theta$,
where $\rho$ is the density at the base of the accreted matter, $h$ is the 
height of the accretion column,
$g$ is the surface gravity, $\mu$ is the shear modulus,
and $\theta$ is the shear angle of the crust.
$\theta$ should be about $10^{-3}$ for neutron stars in low-mass
x-ray binaries to explain the bimodal magnetic field
distribution of binary pulsars({\it{19}}). We expect that $\theta$ for strange stars
in low-mass x-ray binaries is close to this value 
because the stellar crust in both cases are replaced by accretion
material. From ({\it{20}}), 
$\mu \approx 2.5\times 10^{27}\,\,{\rm dyn}\,{\rm cm}^{-2}$, so we obtain
the column density of the accreted matter, $\sigma=\rho h\approx 1.0
\times 10^{10}\theta_{-3}\,\,{\rm g}\,{\rm cm}^{-2}\,$, where $\theta_{-3}
=\theta/10^{-3}\,$. The interval between crust-breaking events can be
written as
\begin{equation}
\tau_1 \approx \frac{\sigma A_p}{\dot {M}}\approx 2.2\times 10^3 
\theta_{-3}L_*^{-5/7}B_*^{-4/7}\,\,{\rm s}\,.
\end{equation}
As the luminosity due to accretion decreases or the magnetic field
decays, $\tau_1$ increases. This time scale is consistent with the 
typical, observed intervals between the HXRBs.
However, the accreted matter may diffuse away from the polar cap area $A_p$
before it builds up enough pressure to break the crust. We now estimate
the diffusion time of the accreted matter across the local magnetic field of
the polar cap $B_s$, which is assumed to be dominated by the multipole field component
and in general much stronger than that of the dipolar field component. 
The transverse velocity resulting from 
collisional diffusion in the presence of a pressure gradient ($\nabla\!P$) 
is approximated by
$v_d\sim 1.3\times 10^2 Z T_8^{-3/2}B_s^{-2}\nabla\!P\,\,{\rm cm}\,{\rm s}^{-1}$
({\it{21}}) where $Z=1$ for hydrogen, $T_8$ is the matter temperature in units of
$10^8\,$K and $B_s$ is the local surface magnetic field. 
In the present case, $\nabla\!P\sim \mu \theta/r_p$. 
So the diffusion velocity $v_d\sim 2\times 10^{-4}T_8^{-3/2}L_*^{-1/7}
B_*^{2/7}(B_s/10^{12}G)^{-2}\,\,{\rm cm}\,{\rm s}^{-1}$. 
Thus, the time scale for the accreted matter to
diffuse over the length $\sim 1\,$km is at least $10^9\,$s, which
is much longer than $\tau_1$.

After the crust has been broken, the accreted matter will fall into the
strange-matter core in a time $\sim 1\,$ms. Subsequently releasing
two kinds of energy: (i) gravitational energy $\sim 2\,$MeV per 
nucleon, which is due to the movement of the accreted matter from the 
stellar surface to the base of the crust and (ii) deconfinement energy $\sim 30\,$MeV
per nucleon, due to the conversion of the accreted matter to more stable 
strange matter ({\it{22}}). The total released energy is $E_{\rm tot}\sim
5.5\times 10^{40}\theta_{-3}L_*^{2/7}B_*^{-4/7}\,{\rm ergs}$. 
Because the total volume of the strange-matter blobs formed during the 
conversion of accreted matter is rather small 
($\sim 10^6\,{\rm cm}^3$), most of $E_{\rm tot}$ will be radiated 
through photons from the surfaces of the blobs. However, a fraction of 
the radiation energy may be absorbed and then reradiated as neutrinos
which pass almost freely through the crust. Thus, it is expected that 
about half of the total released energy may be finally radiated 
in photons which form a fireball of volume $\sim A_pl$ (where $l$ 
is close to the crustal thickness $\sim 10^4\,$cm). Assuming that $\xi$ 
is the ratio of the fireball energy to the total released energy, 
we obtain the fireball energy
\begin{equation}
E_\gamma\sim 2.8\times 10^{40}\xi_{1/2}\theta_{-3}L_*^{2/7}B_*^{-4/7}\,\,
{\rm ergs}\,,
\end{equation}
where $\xi_{1/2}=2\xi$. Let $T_0$ be the initial temperature of the fireball.
By using the expression $\frac{11}{4}aT_0^4A_pl\sim E_\gamma$, we have
$T_0\sim 5.2\times 10^9 \xi_{1/2}^{1/4}\theta_{-3}^{1/4}\,\,{\rm K}$.
The fireball must be contaminated by 
baryons and we can estimate the amount of contamination. The
radiation-dominated outflow begins
when the radiation energy density $u_\gamma=\frac{11}{4}aT_0^4$ is equal 
to the gravitational energy density $u_g=GM\rho/r$, or
$\rho=8.7\times 10^{5}(T_0/10^{10}\,{\rm K})^{4}\,\,{\rm g}
\,{\rm cm}^{-3}$.
>From ({\it{23}}), the column density for the radiation-dominated surface 
layer is given by
$\sigma^{\prime}\approx 3\times 10^8\,\mu_e^{-1/3}(\rho/10^6\,
{\rm g}\,{\rm cm}^{-3})^{4/3}\,\,{\rm g}\,{\rm cm}^{-2}$,
where $\mu_e$ is the mean molecular weight per electron ($\mu_e=1$ for
hydrogen). Therefore, the amount of the baryons
loaded with the fireball is approximated by
$\Delta\!M \approx \sigma^{\prime}A_p
\sim 1.4\times 10^{18}\xi_{1/2}^{4/3}\theta_{-3}^{4/3}
L_*^{2/7}B_*^{-4/7}\,\,{\rm g}$.
Thus, the ratio of the initial fireball energy to the rest energy of the
contaminating baryons is defined as
$\eta\equiv \frac{E_\gamma}{\Delta\!Mc^2}\sim 21 \xi_{1/2}^{-1/3}\theta_{-3}
^{-1/3}$.
The fireball will expand outward because of the large optical depth 
of photon-electron scattering. Because $\Delta M/M_\odot>1.7\times 10^{-16}
(E_\gamma/10^{41}\,{\rm ergs})$({\it{24}}), most of the initial
fireball energy will be converted into the bulk kinetic energy of
the baryons during the expansion. When the optical depth becomes one, 
therefore, the rest radiation energy becomes negligibly small. 
Fortunately, as suggested by M\'esz\'aros and Rees ({\it{25}}),
the expanding shell (having a relativistic factor $\Gamma\sim \eta$) 
will interact with the surrounding medium and its kinetic energy will be
converted into the random energy of the shell by shock waves and finally 
radiated through nonthermal processes in these shock waves. 
The timescale for radiation is approximated by ({\it{25}})
\begin{equation}
\tau_2  \approx  0.1E_{\gamma,40}^{1/3}(\Gamma/10^2)^{-8/3}n_0^{-1/3}\,\,
{\rm s}
\sim  9.0\xi_{1/2}^{11/9}\theta_{-3}^{11/9}L_*^{2/21}B_*^{-4/21}n_0^{-1/3}
\,\,{\rm s}\,,
\end{equation}
where $n_0$ is the interstellar density ($\sim 1\,{\rm cm}^{-3}$). This 
time scale is consistent with the typical, observed durations of 
the HXRBs.

It is widely believed that electrons can be accelerated by shock waves
to very high energy with the minimum Lorentz factor $\gamma_{\rm min}
\sim (m_{p}/m_{e})\Gamma$, assuming that all particles behind the shock waves
have the same energy. If the shock waves can produce approximate equipartition
between the magnetic field energy density and the particle energy density,
then the strength of magnetic field is about 
$B \simeq 0.3\Gamma n_{0}^{1/2} \,\,{\rm G}$({\it{25}}).
The typical synchrotron photon energy emitted by electrons with 
the Lorentz factor $\gamma_{\rm min}$ at the observer's frame is about
\begin{equation}
\epsilon_p \simeq 1\xi_{1/2}^{-4/3}\theta_{-3}^{-4/3}n_{0}^{1/2} \,\,
{\rm KeV} \,.
\end{equation}
It is usually expected that diffusion shock wave acceleration can produce
a power-law electron spectrum, $dN_e/d\gamma \sim \gamma^{-p}, \,
\gamma_{\rm min} \leq \gamma \leq \gamma_{\rm max}$, where the spectral index $p$
is typically between 2 and 3 ({\it{26}}). The synchrotron radiation for electrons
with such a distribution has a spectrum 
for photon energies larger than $\epsilon_p$ with a power-law form with
the photon index of $\alpha=-(p+1)/2$({\it{27}}). So the theoretical value of $\alpha$
may be from about $-1.5$ to $-2.0$. 
This result is consistent with the observations
by OSSE, $\alpha=-(2.0\pm 0.6)$ ({\it{28}}), and the observations by {\em RXTE},
$\alpha\sim -1.3$ ({\it{7}}). The bursting 
spectrum for our radiation model is similar to the spectrum of the 
persistent emission observed in ({\it{7,28}}). Theoretically the radiation processes
at the surface near a magnetic pole of the accreting strange star are 
complicated, and to study these processes is beyond the scope of our paper. 
Some models studying the radiation processes of an accreting 
neutron star with a rather strong magnetic field may give power-law
spectra with an index near $-1.5$({\it{29}}). 
In addition, it should also be noted
that the spectrum does not evolve during the bursting period 
because the spectral index for the electron distribution behind the shock wave
during the shell's expansion is unlikely to be changed ({\it{30}}). 

We can estimate the recovery
time scale as follows. When a burst occurs, the huge radiation pressure
will push the accretion matter outward over a distance $\Delta r \sim
\frac{E_{\gamma}R_{A}v_{r}}{LR_{\rm pulsar}}$, where $L$ is the accretion 
luminosity and $v_{r}$ is the radial velocity of the accreted matter. 
After a burst, the accretion matter will fall back toward the strange star 
over a time
\begin{equation}
\tau_{3} \sim \frac{\Delta r}{v_{r}} \sim 2.4 \times 10^{3}\xi_{1/2}
\theta_{-3}L_*^{-1} \,\,{\rm s} \,.
\end{equation}
which is consistent with the typical, observed recovery time scale({\it{7}}).

We can compare other characteristics determined from
eqs. (1), (2) and (3) with the observations. First,
the {\em RXTE} observations on GRO J1744-28 between 29 January and 26 April,
1996 indicate ({\it{7}}) that the data of the nonbursting flux from this source 
can be approximately fitted with the straight line flux (mcrab)
$=2703.3-23.0D$, where $D$ is the day number in 1996. If this expression 
can be extrapolated to December 1995, the ratio of the persistent
flux on 5 December 1995 to the persistent flux on 30 January 1996 is about
1.6. Because the interval time scale is 
proportional to $L_*^{-5/7}$(eq.1), the ratio of the typical interval time scales
for the HXRBs on 30 January 1996 to that on 5 December 1995 is about 1.4.  
The observations from {\em RXTE} on 30 January 1996 ({\it{7}}) and BATSE on 5 
December 1995 ({\it{2}}) showed this ratio is $\sim 1.38$. Therefore, eq.(1) is 
consistent with the observations. Second, for our model, the bursting 
flux is obtained by dividing eq. (2) by eq. (3), and this flux
is proportional to $L_*^{4/21}$. This means that the bursting flux is 
weakly dependent upon the persistent flux 
in agreement with the observations from BATSE ({\it{2,8}}) and OSSE ({\it{28}}).

The surface radiation in the crust-breaking region during the bursts
should show pulsations amplitude close to that 
of pulsations during the nonbursting periods. This model implication 
agrees with the observations from OSSE ({\it{28}}) and 
{\em RXTE} ({\it{31}}). However, the reults of RXTE indicates that 
the bursting flux seems approximately linearly proportional to the
persistent flux({\it{32}}). 
The more detail discussion on the discrepancies between
experimental results and model predictions is presented elsewhere({\it{33}}).
According to accretion instability models, on the other hand,
during the accretion instability, a great deal of matter falls
onto the surface near a magnetic pole of the pulsar, and subsequently
a great number of gravitational energies are released and an HXRB is 
thus formed. Following this picture, one should detect 
pulsations during the bursts amplitude of which is much larger than that 
of pulsations during the nonbursting periods.

Finally, we want to remark that a similar strange star model is
recently proposed to explain the soft $\gamma$-ray repeaters({\it{34}}).
The key differences between these two models are (i)the crust cracking
of the soft $\gamma$-ray repeater model results from the spin-down 
of the strange star instead
of accretion, (ii)the amount of energy released from these two
mechanisms differ by two order of magnitude and (iii)the strength of
the dipolar magnetic field in these two kind of sources also differ by
two order of magnitude. (iv)The time scales of energy release are difference
by at least one order of magnitude. These differences make the magnetic 
energy density of the soft $\gamma$-ray repeaters stronger than the radiation
energy density therefore the fireball mechanism cannot be developed.

\begin{center}
\bf{References and Notes}
\end{center}

\baselineskip=3.5mm
\begin{description}
\item $1.$ G. Fishman {\em et al.}, IAU Circ. No. {\bf 6272} (1995).
\item $2.$ C. Kouveliotou {\em et al.}, Nature (London) {\bf 379}, 
	    799 (1996).
\item $3.$ M.H. Finger {\em et al.}, Nature (London) {\bf 381}, 
	    291 (1996).
\item $4.$ P. Daumerie, V. Kalogera, F.K. Lamb and D. Psaltis,
	    Nature (London) {\bf 382}, 141 (1996).
\item $5.$ S.J. Sturner and C.D. Dermer, Astrophys. J. {\bf 465},
	    L31 (1996).
\item $6.$ P. Ghosh and F.K. Lamb, Astrophys. J. {\bf 234}, 296 (1979).
\item $7.$ A.B. Giles {\em et al.}, Astrophys. J. {\bf 469}, L25 (1996).
\item $8.$ X-ray bursts are sudden increases in the x-ray flux of x-ray
	    sources, with rise times of $\le$ 1s, and subsequent decay
	    with characteristic times ranging from 10s to a few minutes.
	    They are classified into two classes: type-I and type-II
	    X-ray bursts. Type-I X-ray bursts are characterized by the relatively
	    long burst intervals (hours to days) and the significant
	    spectral softening during the burst decay as compared to 
	    their type-II conterparts. The differences
	    between type-I(II) bursts and the HXRBs is discussed by
	    W.H.G. Lewin, R.E. Rutledge, J.M. Kommers, J. van Paradijs
	    and C. Kouveliotou, Astrophys. J. {\bf 462}, L39 (1996).
\item $9.$ J.K. Cannizzo, Astrophys. J. {\bf 466}, L31 (1996).
\item $10.$ Pure hadronic matter can exist inside compact stars,
	    for example neutron stars in which most stellar masses are in the form of
	    free protons and neutrons. In extremely high density, the 
	    smaller constituents (quarks) inside the protons and neutrons  
	    are no longer confined and a new phase of matter can occur. It is 
	    called quark matter. The possibility of having such new phase of
	    matter in compact stars are discussed by 
	    A.R. Bodmer, Phys. Rev. {\bf 4}, 1601 (1971); E. Witten,
	     Phys. Rev. D {\bf 30}, 272 (1984).
\item $11.$ R.L. Jaffe and E. Farhi, Phys. Rev. D {\bf 30}, 2379 (1984).
\item $12.$ C. Alcock, E. Farhi and A. Olinto, Astrophys. J. {\bf 310},
	     261 (1986); P. Haensel, J.L. Zdunik and R. Schaeffer,
	     Astron. Astrophys. {\bf 160}, 121 (1986); N.K. Glendenning and
	     F. Weber, Astrophys. J. {\bf 400}, 647 (1992).
\item $13.$  Glitches refer to the sudden spin-up of pulsars. Generally, the 
	     new rotation curve will relax back to the original spin-down
	     curve. Such process is called postglitch relaxation. The 
	     physical mechanism of the postglitch relaxation is discussed
	     by D. Pines and M.A. Alpar, Nature (London) {\bf 316}, 27 (1985).
\item $14.$ G. Baym, in {\em Neutron Stars: Theory and Observation}, eds.
	     J. Ventura and D. Pines, 21 (Kluwer, Dordrecht, 1991).
\item $15.$ K.S. Cheng and Z.G. Dai, Phys. Rev. Lett. {\bf 77}, 1210 (1996).
\item $16.$ N.K. Glendenning, Nucl. Phys. B (Proc. Suppl.) {\bf B24}, 110
	     (1991); N.K. Glendenning, Phys. Rev. D {\bf 46}, 1274 (1992);
	     H. Heiselberg, C.J. Pethick and E.F. Staubo, Phys. Rev. Lett.
	     {\bf 70}, 1355 (1993); V.R. Pandharipande and E.F. Staubo, 
	     in {\em Proc. 2nd International Conf. of Physics and 
	     Astrophysics of Aurk-Qluon Plasma} eds. B. Sinha, Y.P. Viyogi 
	     and S. Raha, (World Scientific, 1994).
\item $17.$ M. Prakash, J.R. Cooke and J.M. Lattimer, Phys. Rev. D {\bf 52},
	     661 (1995).
\item $18.$ R.R. Caldwell and J.L. Friedman, Phys. Lett. B {\bf 264}, 
	     143 (1991);W.Kluzniak, Astron. Astrophys. {\bf 286}, L17 (1994).
\item $19.$ K.S. Cheng and Z.G. Dai, Astrophys. J. {\bf 476}, L39 (1997).
\item $20.$ G. Baym and D. Pines, Ann. Phys. {\bf 66}, 816 (1971).
\item $21.$ L. Spitzer, {\em Physics of Fully Ionized Gases}, 42 (2d ed.;
	     New York: Interscience, 1962).
\item $22.$ G. Baym, E.W. Kolb, L. McLerran, T.P. Walker and J.L. Jaffe,
	     Phys. Lett. B {\bf 160}, 181 (1985).
\item $23.$ P. M\'esz\'aros and M. Rees, Astrophys. J. {\bf 397}, 570 (1992).
\item $24.$ A. Shemi and T. Piran, Astrophys. J. {\bf 365}, L55 (1990).
\item $25.$ P. M\'esz\'aros and M. Rees, Astrophys. J. {\bf 405}, 278 (1993).
\item $26.$ R. Blandford and D. Eichler, Phys. Rep. {\bf 154}, 1 (1987).
\item $27.$ K.S. Cheng and D.M. Wei, Mon. Not. R. Astron. Soc. {\bf 283}, L133 
	     (1996).
\item $28.$ M.S. Strickman {\em et al.}, Astrophys. J. {\bf 464}, L131 (1996).
\item $29.$ P. M\'esz\'aros, {\em High Energy Radiation from Magnetized 
	     Neutron Stars} (Univ. Chicago Press, 1992).
\item $30.$ E. Waxman, Astrophys. J., in press (1997); Nature (London),
	     submitted (1997).
\item $31.$ M.J. Stark, A. Baykai, T. Strohmayer and J. J.H. Swank,
	     Astrophys. J. {\bf 470}, L109 (1996).
\item $32.$ K.Jahoda, M.J. Stark, T.E. Strohmayer and W.Zhang, in the
	     Proceedings of the Symposium "The Active X-ray Sky",Rome, 
	     astro-ph/9712340(1997).
\item $33.$ K.S.Cheng, Z.G.Dai, and T.Lu, Int. J. Mod. Phys. D., in press(1998).
\item $34.$ K.S.Cheng and Z.G.Dai, Phys. Rev. Lett., {\bf80},18 (1998).
\end{description}

\end{document}